\documentclass[aps,prl,twocolumn,superscriptaddress,showpacs,longbibliography]{revtex4-2}
\usepackage{braket}
\usepackage{graphicx}
\usepackage{latexsym}
\usepackage{amssymb}
\usepackage{amsmath}
\usepackage{amsfonts}
\usepackage{bm}
\usepackage{multirow}
\usepackage{color}
\usepackage{xcolor}
\usepackage{calrsfs}
\usepackage{dutchcal}

\usepackage[colorlinks=true, citecolor={blue!80!black}, urlcolor={blue!50!black}, linkcolor = {blue!80!black},hypertexnames=false]{hyperref}

\usepackage[percent]{overpic}
\usepackage{tabularx}

\newcommand{\be}{\begin{equation}}
\newcommand{\ee}{\end{equation}}
\newcommand{\bea}{\begin{eqnarray}}
\newcommand{\eea}{\end{eqnarray}}

\begin{document}

\title{Spin Seebeck effect of interacting spinons}

\author{Ren-Bo Wang}
\affiliation{French American Center for Theoretical Science, CNRS, KITP, Santa Barbara, California 93106-4030, USA}
\affiliation{Kavli Institute for Theoretical Physics, University of California, Santa Barbara, California 93106-4030, USA}

\author{Naveen Nishad}
\affiliation{Physics Department, Technion, 32000 Haifa, Israel} 
\affiliation{The Helen Diller Quantum Center, Technion, 32000 Haifa, Israel}
\author{Anna Keselman}
\affiliation{Physics Department, Technion, 32000 Haifa, Israel} 
\affiliation{The Helen Diller Quantum Center, Technion, 32000 Haifa, Israel}
\author{Leon Balents}
\affiliation{Kavli Institute for Theoretical Physics, University of California, Santa Barbara, California 93106-4030, USA}
\affiliation{Canadian Institute for Advanced Research, Toronto,  Ontario, Canada}
\author{Oleg A. Starykh}
\affiliation{Department of Physics and Astronomy, University of Utah,  Salt Lake City, Utah 84112, USA}

\begin{abstract}
We present the theory of the longitudinal spin Seebeck effect between a Heisenberg spin-$1/2$ chain and a conductor. The effect consists of the generation of a spin current across the spin chain-conductor interface in response to the temperature difference between the two systems. In this setup, the current is given by the convolution of the local spin susceptibilities of the spin chain and the conductor. We find the spin current to be fully controlled, both in the magnitude and the sign, by the backscattering interaction between spinons, fractionalized spin excitations of the Heisenberg chain. In particular, it vanishes when the spinons form a non-interacting spinon gas. Our analytical results for the local spin susceptibility at the open end of the spin chain are in excellent agreement with numerical DMRG simulations.
\end{abstract}

\maketitle

\emph{Introduction.} The nearest-neighbor antiferromagnetic spin-$1/2$ chain~\cite{Bethe1931} represents the simplest and best understood example of a quantum spin liquid, with long-ranged algebraic correlations between constituent spin-$1/2$ magnetic moments and a robust multi-particle continuum of fractionalized elementary excitations called spinons.  Over its venerable history, the spin-1/2 chain has proved a remarkably rich source of new physics, slowly revealing yet more intricate structure over time \cite{Stone2003,Mourigal2013,Lorenz2017}. Here we uncover another novel property of the chain: the local dynamical susceptibility for the spin at the end of the chain displays an intricate structure as a function of frequency and applied field which is strongly influenced by interactions between spinons.   In particular, we demonstrate a ``peak-gap-hump" feature occurs as a function of frequency whose strength and character is entirely determined by spinon-spinon interactions, which can be tuned by varying the second neighbor exchange of the chain.  

While new insights into the $S=1/2$ chain are significant purely based on the fundamental nature of the problem, our result acquires additional importance in the context of spintronics \cite{Han2018,SSE-review,Takahashi2016,nematic2019,triplon2021,apl2022}.  Recent experiments report an observation of the spin Seebeck effect (SSE)
in a Sr$_2$CuO$_3$/Pt hybrid structure \cite{Hirobe2017,Hirobe2018}. 
Sr$_2$CuO$_3$ is a quasi-one-dimensional magnetic insulator made of Heisenberg spin-$1/2$ chains with large exchange interaction $J \approx 2200$ K between nearest spins in a chain and weak interchain exchange $J_{\rm inter} \approx 5$ K between spins on the nearest chains. The SSE consists of the generation of a spin current across the spin chain-conductor interface in response to the temperature difference between the two systems. The spin current injected from the spin chain into Pt metal is converted into a measurable voltage via the inverse spin Hall effect \cite{SSE-review}. Specifically, the experiment observed a {\em negative} spin Seebeck effect in Sr$_2$CuO$_3$/Pt structure in a wide temperature range $5~{\rm K} \sim J_{\rm inter} < T \leq 200~{\rm K} \ll J$ where correlations between spins from the same chain are strong but the inter-chain correlations are absent, i.e.\ the chains can be viewed as decoupled from each other \cite{Kojima1997}. The negative sign of the spin current refers to it being opposite to that of the more conventional ferromagnet/metal interface under similar experimental conditions \cite{Hirobe2017,SSE-review}. Given this setting, the spin current measured in the experiment must be carried by fractionalized spinons and, as such, represents an interesting novel probe of the one-dimensional spin liquid.  


\begin{figure}[h]
\centering
\begin{overpic}[width=0.8\columnwidth]{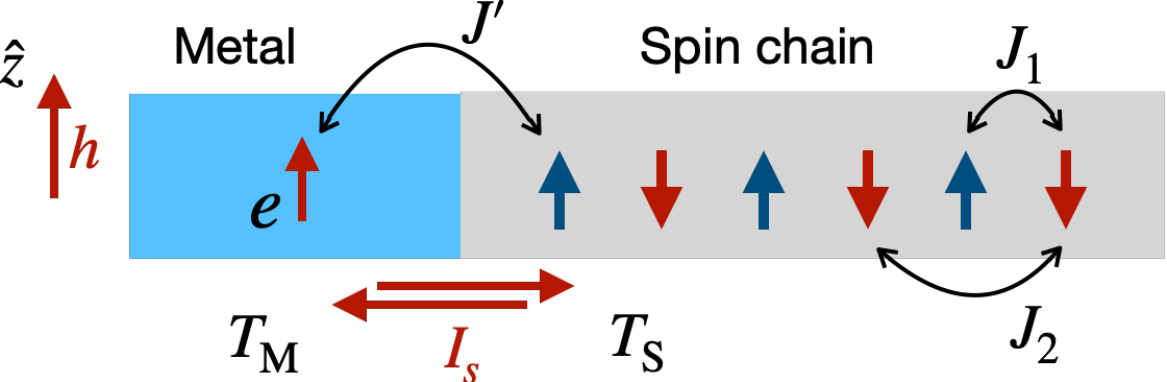}
 \put (-5,12) {\footnotesize{(a)}} \end{overpic}
\begin{overpic}[width=0.8\columnwidth]{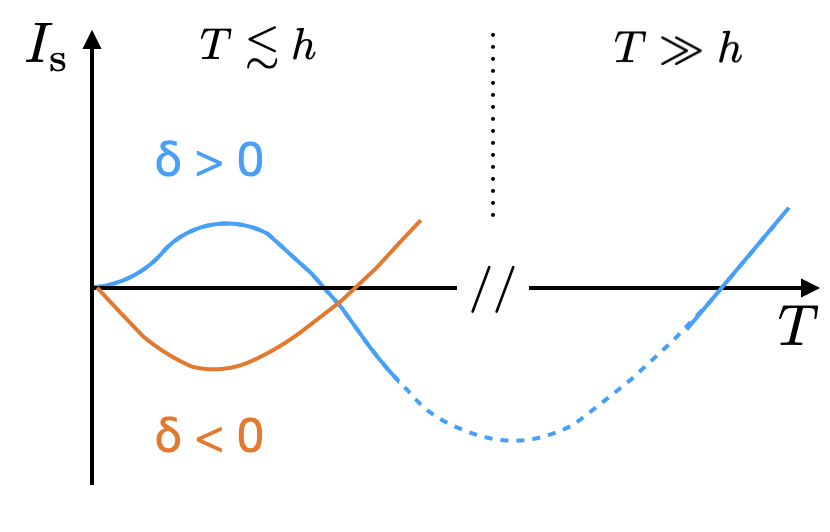}
 \put (-5,32) {\footnotesize{(b)}} \end{overpic}
    \caption{(a) Schematic setup: a quasi-1D magnetic insulator is coupled to a non-magnetic metal via exchange interaction $J'$. In the presence of an external magnetic field, a temperature gradient across the interface gives rise to a spin current $I_s$ between the insulator and the metal. The temperatures of the insulator (metal) are denoted by $T_{\rm S}$ ($T_{\rm M}$). (b) Sketch of $I_{\rm s}(T)$ behavior for $\delta > 0$ and $\delta < 0$ according to field-theoretic calculations for $ h \ll J$.}
    \label{fig:setup}
\end{figure}

We show that, indeed, the spin current follows from the dynamical local susceptibility at the end of the open spin chain.  Our results for the latter imply that the low-temperature asymptotic of the spin current carried by spinons is completely determined by the interaction between them.  In particular, the spin current vanishes when spinon quasiparticles form a non-interacting spinon gas.

\underline{The setup.}
A schematic figure of spin current injection is shown in Fig.~\ref{fig:setup}(a). 
The spin chain is described by the semi-infinite spin-1/2 Heisenberg model with interactions $J_1$ ($J_2$) between the nearest (next-nearest) spins localized on the sites of a one-dimensional lattice,
\be
H_{\rm S} = \sum_{\ell=1,2} \sum_{n=1}^\infty J_\ell {\bf S}_n \cdot {\bf S}_{n+\ell} - h S^z_n .
\label{eq9}
\ee
The metal $H_{\rm M}$ is described by a non-interacting electron gas. 
The interface between the spin chain and the metal is modeled by the exchange interaction between the spin ${\bf S}_1$ on the first site of the spin chain and the conduction-electron spin density ${\mathbf s}_0$ at the point of contact between the chain and the metal,
\be
H' = J' {\mathbf s}_0 \cdot {\bf S}_1.
\label{eq1}
\ee
The assumption of the spin conservation made in \eqref{eq1} is standard in SSE studies \cite{SSE-review}. While the magnitude of the SSE signal does depend on the purity of the spin chain material and the oxidation level of the interface, the negative sign of the spin current is robust to these experimental perturbations \cite{Hirobe2018}.

The spin current per chain, $I_{\rm s} = \langle d\, {s}^z_0/dt \rangle$, is calculated in the linear response theory (the second-order perturbation theory) with respect to the weak interface exchange $J'$ \cite{Jauho1994,SSE-review,Hirobe2017,Sato2024},
\be
I_{\rm s} = \frac{J'^2 \Delta T}{4 T^2} \int_{-\infty}^\infty \frac{d \omega}{2\pi} \frac{\omega}{\sinh^2(\frac{\omega}{2T})} \, {\rm Im}\chi^\pm_{\rm M}(\omega) \, {\rm Im}\chi^\pm_{\rm S}(\omega)  .
\label{eq2}
\ee
The factor $\omega/\sinh^2(\omega/2T)$ comes from expanding the difference of the Bose functions describing the thermal population of spin excitations in the metal and spin chain under the assumption that the temperature difference between the spin chain and metal is small, $T_{\rm M} - T_{\rm S} = \Delta T \ll T = (T_{\rm M} + T_{\rm S})/2$.
Here $\chi^\pm_{\rm M/S}(\omega)$ is the local transverse spin susceptibility of the metal/spin chain {\em at the interface}, respectively. Namely,
\be
\chi^\pm_{\rm S}(\omega) = \int_{-\infty}^\infty dt \, e^{i \omega t} (-i \theta(t)) \langle [S_1^+(t), S_1^-(0)]\rangle_0 ,
\label{eq3}
\ee
is the transverse spin susceptibility of the first (leftmost) spin ${\bf S}_1$ of the {\em semi-infinite} spin chain. The chain is {\em open}, that is subject to the boundary condition ${\bf S}_0(t)=0$ for all times. 
The brackets $\langle \cdots \rangle_0$ denote the average with respect to the decoupled metal/spin chain system (for which the Hamiltonian is $H_0 = H_{\rm S} + H_{\rm M}$ without $H'$ above). This crucial technical calculation is described in the Supplemental Material~\cite{SM}.

\underline{Local susceptibility.} The dynamical response of the spin chain at frequency $\omega \ll J_1$ is described by the right/left moving currents ${\bf J}_{R/L}(x,t)$ \cite{gogolin2004bos,KBS2020,RBWang2022}. The open boundary condition at $x=0$ ``ties" the left-moving spin current to the right-moving one, ${\bf J}_L(x,t)={\bf J}_R(-x,t)$~\cite{Fabrizio1995,affleck1992}, resulting in the chiral form of the low-energy Hamiltonian for the
semi-infinite chain
\begin{align}
\label{eq:Hs}
H_{\rm S} 
&= \int_{-\infty}^\infty dx \, \Big(\frac{2\pi v}{3} {\mathbf J}_R(x)\cdot {\mathbf J}_R(x) - \frac{g_{\rm bs}}{2} {\mathbf J}_R(x)\cdot {\mathbf J}_R(-x) \nonumber\\
&- h { J}^z_R(x)\Big).
\end{align}
The initial value of the backscattering interaction $g_{\rm bs}$ between spinons is determined by the ratio of exchanges $J_2/J_1$, reaching zero at $J_{2,c} \approx 0.24 J_1$ \cite{eggert1996,KBS2020}.
The calculation described in \cite{SM}, based on the hydrodynamic approximation developed in \cite{KBS2020,RBWang2022}, allow us to derive the simple analytical result for the transverse dynamic susceptibility of the open spin-1/2 chain, 
\begin{align}
\label{eq7}
  {\rm Im}\,\chi^\pm_{\rm S}(\omega) = - \frac{\chi \,\omega}{\tilde{v}} 
  \begin{cases}
 \sqrt{\frac{h'-\omega}{h-\omega}}, & \, \omega < -h_{\rm min}; ({\rm a})\\
  f_{\rm L}(\omega) \,\sqrt{\frac{h'-\omega}{h-\omega}}, & \, -h_{\rm min} <\omega < h_{\rm min}; ({\rm b})\\
  0, & \, h_{\rm min} < \omega < h_{\rm max}; ({\rm c}) \\
  \sqrt{\frac{\omega-h'}{\omega-h}}, & \omega > h_{\rm max} ; ({\rm d})
  \end{cases}
\end{align}
where $\chi=\chi_0/(1-\delta)$, $\chi_0 = 1/(2\pi v)$ and $\tilde{v}= v\sqrt{1-\delta^2}$ are the renormalized static susceptibility and spinon velocity, and $h' = h(1+\delta)/(1-\delta)$ is the precession frequency of the spin-current mode. Here $\delta = g_{\rm bs}/(4\pi v)$ is the key dimensionless interaction parameter of the theory. In the renormalization group (RG) terminology, $g_{\rm bs}$ is an energy-dependent running parameter, and here by $\delta$ we denote its value at the Zeeman energy, $\delta = \delta(h)$. 
We also defined $h_{\rm min} = {\rm min}(h, h')$ and $h_{\rm max} = {\rm max}(h, h')$. 
In this way, $h_{\rm min} \, (h_{\rm max}) = h \, (h')$ for $\delta > 0$, correspondingly, while for negative $\delta < 0$ they switch, $h_{\rm min} \, (h_{\rm max}) = h' \, (h)$. As written, Eq.~\eqref{eq7} holds for any sign of $\delta$.

The spectral gap  (i.e.\ ${\rm Im}\,\chi^\pm_{{\rm S}}=0$) within the interval (c) $h_{\rm min} < \omega < h_{\rm max}$ is a characteristic feature of the interacting spinon liquid \cite{KBS2020}. 
It is a direct consequence of the finite energy splitting between the Larmor and spin-current processing modes at small momenta, as discussed in \cite{KBS2020}. Note that \eqref{eq7} diverges as $\omega \to h$, reflecting the diverging density of states associated with the Larmor mode. At the same time, for $\omega \to h'$ the susceptibility vanishes in a square-root fashion, reflecting the vanishing spectral weight $\sim k^2$ of the spin-current mode for $k\to 0$. 

Finally, the function $f_{\rm L}(\omega) = (\omega/\tilde{v})^{-2\delta}/\Gamma(2-2\delta)$ in \eqref{eq7} accounts for the Luttinger liquid renormalization of the spectral weight at frequencies {\em below} the Zeeman one \cite{affleckoshikawa1999}. As a result, it is localized within the frequency interval $|\omega| < h_{\rm min}$, which we denote as interval ${\rm (b)}$ in \eqref{eq7}, while outside of it $f_{\rm L} = 1$.

Observe also that the non-interacting result ${\rm Im}\,\chi^\pm_{\rm S}(\omega) \sim \omega$ is recovered by \eqref{eq7} in the $\delta \to 0$ limit which describes a non-interacting spinon gas that is realized at the endpoint of the critical Luttinger phase in the $J_1$-$J_2$ Heisenberg chain at $J_{2,{\rm c}} \approx 0.24 J_1$ \cite{KBS2020}.

For the metal ${\rm Im}\chi^\pm_{\rm M}(\omega) = -\pi {\cal D}_F^2 \, \omega$ for small frequency (relative to the Fermi energy), and ${\cal D}_F$ is the density of states at the Fermi level.


\underline{Spinon spin current.} Plugging in the expressions for the local susceptibility of the metal and the spin chain discussed above into Eq.~\eqref{eq2}, we obtain 
the spinon current $I_{\rm s} = C \, j \,\Delta T$, where $C = J'^2 {\cal D}_F^2 \chi/(2 \tilde{v})$ and 
\be
j(\delta, h, T) = \int_{-\infty}^\infty \, d\omega \frac{\omega^2 (-\tilde{v} \, {\rm Im}\chi^\pm_{\rm S}(\omega)/\chi)}{4 T^2 \sinh^2(\frac{\omega}{2T})} = \sum_{{\rm n} = {\rm a,b,d}}\tilde{j}_{\rm n}.
\label{eq8b}
\ee
Here $\tilde{j}_{\rm n}$ describes contributions from the corresponding frequency intervals in \eqref{eq7}.

Our key result immediately follows from \eqref{eq8b}: for $\delta=0$ the integrand is an {\em odd} function of $\omega$ and hence $I_{\rm s}(\delta=0)=0$. Non-interacting spinons do not have an SSE. 

The low-$T$ asymptotic for $T \ll h$ is determined by $\tilde{j}_{\rm b}$, the contribution from the ${\rm (b)}$-interval  $|\omega| < h_{\rm min}$, which contains the lowest frequencies. 
It is obtained by expanding the square root in small $\omega \sim T \ll h$ and approximating integration limits of the interval ${\rm n = b}$ by $\pm \infty$. This gives
\be
\tilde{j}_{\rm b} \approx \frac{2\delta}{\sqrt{1-\delta^2}} \frac{(2 T)^{3-2\delta}}{\tilde{v}^{-2\delta} \, h} \int_0^\infty \frac{dx \, x^{4-2\delta}}{\sinh^2(x)},
\label{eq10}
\ee
where the last integral is approximately $\pi^4/30-4.29 \delta$ for the relevant $\delta \sim 0.1$. Crucially, the {\em sign} of the low-$T$ asymptotic is determined by the sign of $\delta$. As already noted above, $\tilde{j}_{\rm b}$, as well as the total $I_{\rm s}$, vanishes for $\delta=0$.
(Note that for $\delta < 0$, a finite magnetic field is required to maintain the Luttinger liquid ground state \cite{hikihara2010}. This condition is fulfilled in our study.)

As $T$ increases, the effective range of $\omega$ increases with it, and for 
$T_{\rm sign}$ of order $h$ (numerically, we find it to be approximately $0.3 h$), the negative frequency contribution $\tilde{j}_{\rm a}$ in \eqref{eq8b} overcomes those from $\tilde{j}_{\rm b}$ and $\tilde{j}_{\rm d}$. As a result, for $T > T_{\rm sign}$, the net spin current $j$ {\em changes sign} for either sign of $\delta$. The high-temperature limit, $\omega, h \ll T$, of \eqref{eq2} is interesting too. 
Equations \eqref{eq3} and \eqref{eq2} show that $I_{\rm s} \propto -\int d\omega \,{\rm Im}\chi^\pm_{\rm S}(\omega) \propto m/\chi = h$, where $m$ is the magnetization (see Eq.\ (\ref{end:highT}) in \cite{SM} for details). 
That is, $I_{\rm s}$ {\em must be positive} at sufficiently high $T$. Together with the low-$T$ asymptotics, this implies a highly non-monotonic dependence of $I_{\rm s}$ on $T$, with {\em two} changes of the sign of the spin current for $\delta >0$, as illustrated in Fig.~\ref{fig:setup}(b). 


A brief comment is in order: Eq.~\eqref{eq7} is derived at zero temperature $T=0$ and used in \eqref{eq8b} and \eqref{eq10} to calculate the low-$T$ behavior of the spin current. 
The appropriateness of this procedure follows from noting that (i) $\chi_{\rm S}^\pm(k,\omega)$ (Eq.\ (\ref{4.57}) in \cite{SM}), from which \eqref{eq7} follows, is obtained via the perturbation theory formulated entirely in terms of spin currents ${\bf J}_{R/L}$ (this point is made explicit in the SM of \cite{KBS2020}), and (ii) finite-$T$ Green's functions of spin currents depend on $T$ only via the $T$-dependence of the Matsubara frequency \cite{sachdev1994}. As a result, the leading $T$-dependence of the retarded susceptibility $\chi_{\rm S}^\pm(\omega)$ is via that of $\delta(T)$. In the presence of the magnetic field $h$, that dependence is weak for $T < h$, when the RG length associated with the magnetic field, $1/h$, is shorter than the one due to the temperature $1/T$ \cite{affleckoshikawa1999,affleck-takahashi,giamarchi2003}.
This is supported by our finite-$T$ numerical simulations of the local susceptibility presented in \cite{SM}, Fig.~\ref{fig:Susceptibility_vs_beta}.

\underline{Comments on the theory proposed in \cite{Hirobe2017}.} In their theoretical modeling of the spinon spin current, Ref.\ \cite{Hirobe2017} has ignored the need for the open boundary condition for the spin chain, which is fully implemented in our Eq.\ \eqref{eq:Hs}, and completely neglected the contribution of the chiral currents ${\bf J}_{R/L}$ to the boundary spin susceptibility, which in fact, is the only contribution to $\chi^\pm_{\rm S}(\omega)$. In place of $\chi^\pm_{\rm S}$ \eqref{eq7}, \cite{Hirobe2017} used the staggered susceptibility (which vanishes near the open boundary), which, moreover, was extended to include excitations with energies of order  $J_1$ so as to capture the crucial even-in-$\omega$ piece of $\chi$. The latter step has no theoretical justification within the low-energy field-theoretical description of the spin chain. 

It is worth reiterating here that in our theory, the crucial non-linearity of $\chi^\pm_{\rm S}(\omega)$ is due to the spectral gap, frequency interval (c) in Eq.\ \eqref{eq7}, which is the intrinsic low-energy feature of the {\em interacting} spinon liquid.

\underline{Numerical simulations.}
We next compare our analytical results to numerical matrix product state (MPS)-based simulations. To obtain the local susceptibility~\eqref{eq3}, we first obtain the ground state of the system using density matrix renormalization group (DMRG)~\cite{White1992,Schollwock2011}. We then perform time evolution of the quenched state with a flipped spin at the boundary, up to times $t_{\rm max} = 200 J^{-1}$ using time evolving block decimation (TEBD)~\cite{Vidal2004,Paeckel2019}. Before carrying out a Fourier transform of the resulting correlations we use linear prediction to extrapolate the correlations up to times $2t_{\rm max}$ followed by windowing using a Gaussian with $\sigma=0.9t_{\rm max}$~\cite{White2008}.
Our analysis is done on systems of length $N = 240$ sites.
All numerical calculations were carried out using the
ITensor library~\cite{itensor}.

\begin{figure}
    \centering
    \begin{overpic}[width=0.85\linewidth]{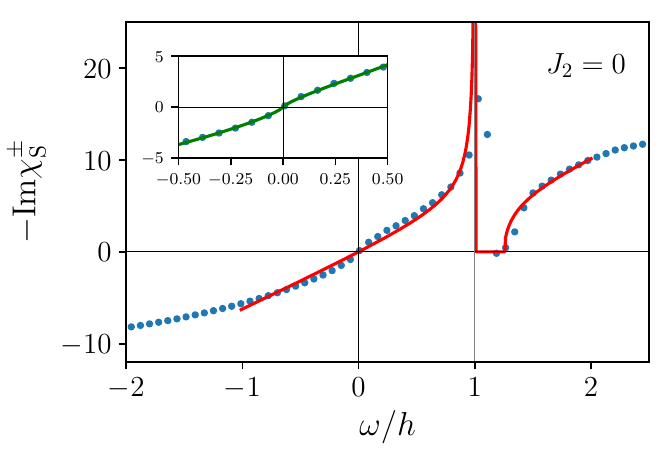}
    \put (-5,62) {\footnotesize{(a)}} \end{overpic}
    \begin{overpic}[width=0.85\linewidth]{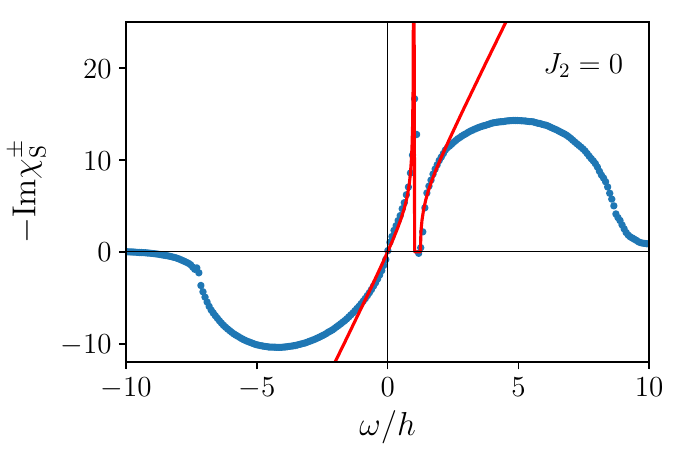}
    \put (-5,62) {\footnotesize{(b)}} \end{overpic}
    \begin{overpic}[width=0.85\linewidth]{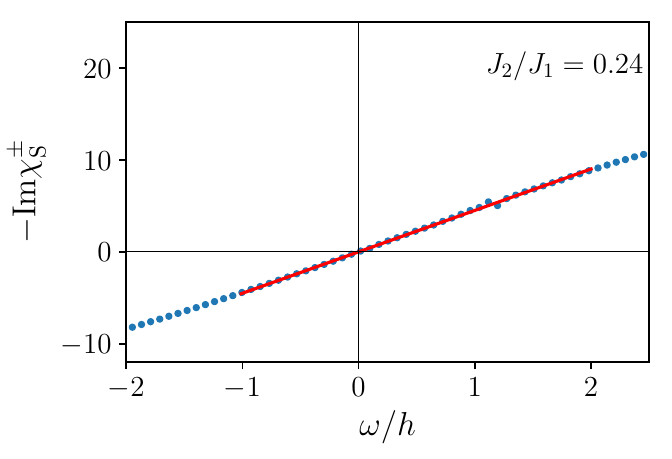}
    \put (-5,62) {\footnotesize{(c)}} \end{overpic}
    \begin{overpic}[width=0.85\linewidth]{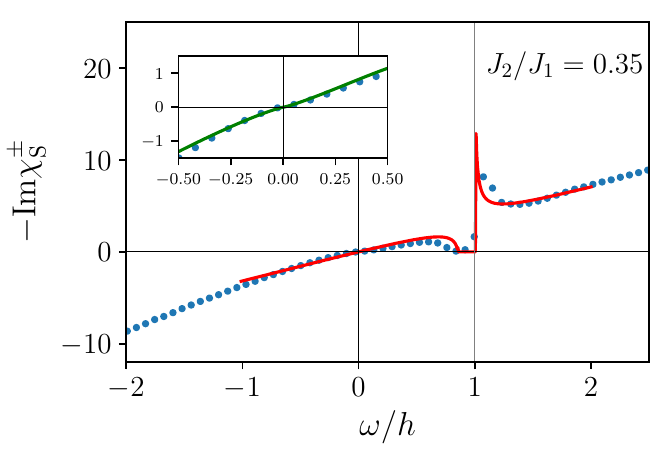}
    \put (-5,62) {\footnotesize{(d)}} \end{overpic}
    \caption{Local susceptibility obtained numerically for $h/J_1=0.2$ and three different values of $J_2/J_1=0,0.24,0.35$ in panels (a), (c), (d), respectively.     Red solid lines are fits to the analytical form \eqref{eq7} with $f_{\rm L}(\omega)=1$,  yielding $\delta=0.11,0,-0.08$, respectively. Insets in (a) and (d) show zoom in on the low frequencies region, where the green solid line is a fit to the form in \eqref{eq7}(b) with the full expression for $f_{\rm L}(\omega)$ using $\delta$ obtained above. (b) is the zoom out of panel (a).}
    \label{fig:local_chi_numerics}
\end{figure}

Results for the local susceptibility at $T=0$ as well as fits to the behavior expected from the low energy theory are plotted in Fig.~\ref{fig:local_chi_numerics}. 
We observe an excellent agreement in the low frequency regime both for $J_2\leq J_{2,c}$ and $J_2>J_{2,c}$, corresponding to $\delta\geq 0$ and $\delta<0$, respectively. 
Deviations from the low energy form can be seen at frequencies $\omega\lesssim J_1$, as expected, due to the finite bandwidth (see \cite{SM} for further discussion). 




\begin{figure}
    \centering
    \includegraphics[width=\columnwidth]{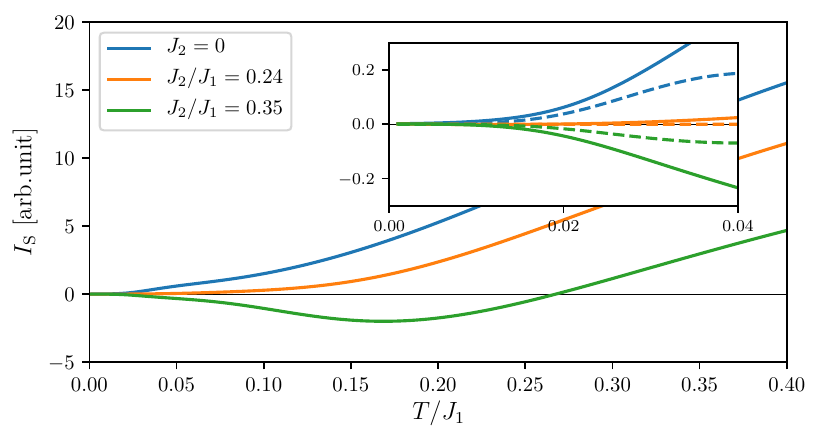}
    \caption{Integrated current obtained using the ground state susceptibilities for $h/J_1=0.2$ and three different values of $J_2/J_1=0,0.24,0.35$. Inset shows zoom-in on the low temperature regime, with dashed lines corresponding to the current expected from the field theory expression Eq.~\eqref{eq8b} with $\delta$ obtained from the fits in Fig.~\ref{fig:local_chi_numerics}.}
    \label{fig:current_numerics}
\end{figure}

In Fig.~\ref{fig:current_numerics} we plot the spin current, Eq.~\eqref{eq2}, obtained using the ground state susceptibilities calculated above, which is appropriate for $T\ll J$ limit as discussed previously. 
We have also verified numerically using finite temperature MPS simulations \cite{White2009, Stoudenmire2010} that the gap in the susceptibility persists up to temperatures of order the Zeeman energy $h$ (see Fig.\ \ref{fig:Susceptibility_vs_beta} in \cite{SM}).
As can be seen in the inset, the behavior of the current at low temperatures $T\ll h,J_1$ is as expected from the low energy theory (see Fig.\ \ref{fig:chi}(c,d)), while at larger temperatures $T\lesssim J_1$ deviations from this behavior can be observed due the finite bandwidth of the chain {\em and} the fact that the magnetic field used in the simulation, $h = 0.2 J_1$, is not quite in the field theory regime of $h \ll J_1$. Fig.~\ref{fig:local_chi_numerics}(b) shows that numerical ${\rm Im}\chi^\pm_{\rm S}(\omega)$ (blue curve) that is used in \eqref{eq2} deviates from the field theory form (red curve) already for $\omega/h \sim 3$. As a result, numerically obtained spin current for the $J_2=0$ (blue) curve in Fig.~\ref{fig:current_numerics} does not change sign as $T$ increases beyond $0.3 h$, 
in disagreement with low-energy predictions shown in Fig.\ \ref{fig:chi}(c) and Fig.~\ref{fig:setup}(b). 

The crucial role of the finite bandwidth and curvature of the spinon dispersion is best illustrated by considering the chain tuned to the critical $J_{2,c} = 0.24 J_1$ which at low energies corresponds to the non-interacting field theory with $\delta=0$, see Fig.~\ref{fig:local_chi_numerics}(c). In this case the field theory predicts $I_{\rm s} = 0$. The numerical result is shown by the orange curve in Fig.~\ref{fig:current_numerics}. At very low $T$ the spin current is indeed very strongly suppressed relative to the two other cases with $\delta \neq 0$. However, at higher temperatures $T \approx 0.2 J_1$ and above, the spin current becomes comparable to the two other cases. This is understood as being due to the non-linear dispersion of the spinons.

\underline{Discussion.} Our study of the spinon spin current represents a novel extension of quantum spintronics research \cite{SSE-review,Takei2024,Sato2024}. We uncover a dramatic sensitivity of the spin current to the interactions between spinons and its inherently non-monotonic temperature dependence, Figs.\ \ref{fig:setup}(b) and \ref{fig:current_numerics}.
Intriguingly, Ref.~\cite{Hirobe2017} reported {\em two} sign changes of the measured spin current with decreasing $T$: the current changed from positive to negative at about 200\,K and then it turned back from negative to positive at about 5\,K. The authors \cite{Hirobe2017} have attributed the 5\,K change of the spin current sign to the development of the long-range antiferromagnetic order. Our theory offers a compelling alternative interpretation of this unusual behavior - it is an intrinsic feature of the critical system of {\em interacting spinons}. 
It is worth noting that the very large value of the exchange interaction $J=J_1$ in this material, about $2200$\,K, makes the `field theory' regime $h \ll J_1$ easily satisfiable in the experiment. Moreover, our theory also explains why the spin current must be positive at sufficiently high $T$ by tying it to the sign of magnetization, as in Eq.\ (\ref{end:highT}). 

On the theory side, a better understanding of the interplay between inter-spinon interactions and the non-linearity of their dispersion \cite{imambekov2012} is required. Our numerical findings, Fig. \ref{fig:current_numerics}, show that non-linear dispersion and finite bandwidth dominate over the described field-theoretical limit at higher $T$. We leave this interesting topic to future investigations.


\begin{acknowledgments}
R.B.W.\ was funded by the European Research Council (ERC) under the European Union's Horizon 2020 research and innovation program (Grant Agreement No.\ 853116, acronym TRANSPORT).  
A.K.\ acknowledges funding by the Israeli Council for Higher Education support program and by the Israel Science Foundation (Grant No.\ 2443/22). 
L.B.\ is supported by the NSF CMMT program under Grant No.\  DMR-2419871, and the Simons Collaboration on Ultra-Quantum Matter, which is a grant from the Simons Foundation (Grant No.\ 651440).  
O.A.S.\ was supported by the NSF CMMT Grant No.\ DMR-1928919. 
This research was supported in part by grant NSF PHY-2309135 to the Kavli Institute for Theoretical Physics (KITP).
\end{acknowledgments}

\bibliography{spinon-current-refs}

\clearpage

\setcounter{page}{1}
\setcounter{equation}{0}
\setcounter{figure}{0}
\setcounter{secnumdepth}{3}
\renewcommand{\theequation}{S\arabic{equation}}
\renewcommand{\thefigure}{S\arabic{figure}}

\begin{center}
{\Large\bfseries Supplementary Material for ``Spin Seebeck effect of interacting spinons" }
\end{center}

The Supplemental Material contains
the summary of field-theoretical predic-
tions for the end-chain susceptibility and asymptotic behavior
of the spin current, detailed derivation of the susceptibility,
explanation of the low-energy Luttinger physics corrections,
derivation of the dynamical susceptibility of a noninteracting
fermion chain, and numerical simulations of the end-chain sus-
ceptibility at finite temperatures.

\section{Summary of the field-theoretical predictions for the end-chain spin susceptibility and the spin current}

\begin{figure}
    \centering
    \begin{overpic}[width=0.9\linewidth]{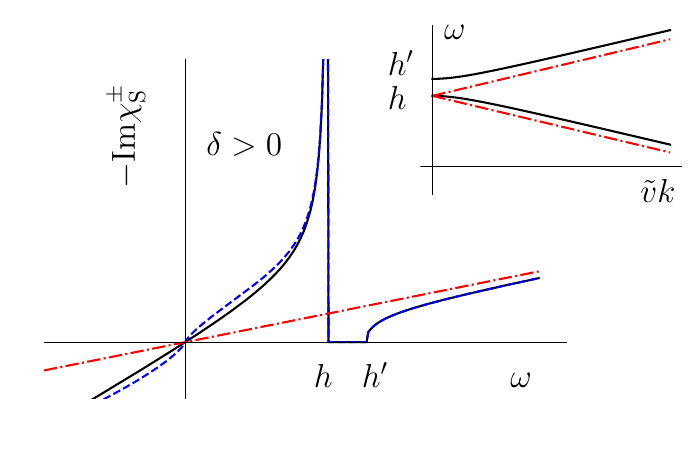}
    \put (-5,62) {\footnotesize{(a)}} \end{overpic}
    \begin{overpic}[width=0.9\linewidth]{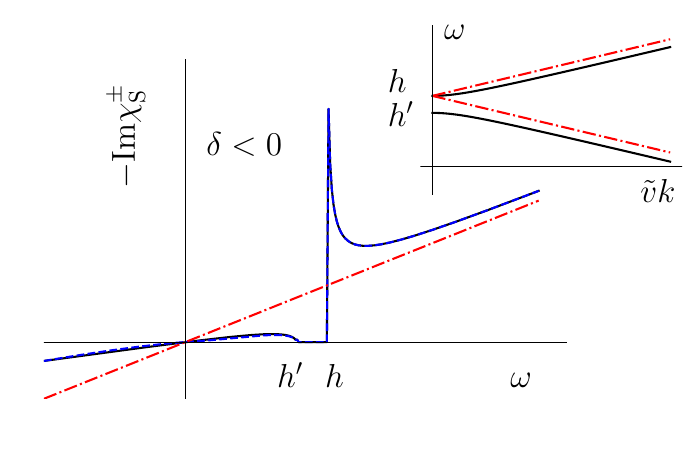}
    \put (-5,62) {\footnotesize{(b)}} \end{overpic}
    \begin{overpic}[width=0.8\linewidth]{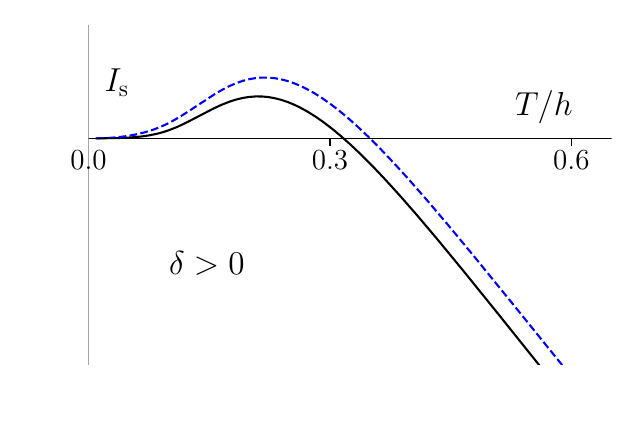}
    \put (-5,62) {\footnotesize{(c)}} \end{overpic}
    \begin{overpic}[width=0.8\linewidth]{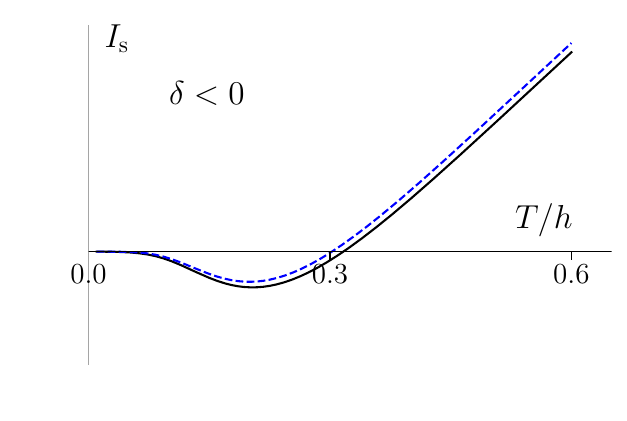}
    \put (-5,62) {\footnotesize{(d)}} \end{overpic}
       \caption{(a,b) Local susceptibility at the end of the spin chain, Eq.\ (6), for $\delta > 0$ in (a) and $\delta < 0$ in (b). Here $h' = h(1+\delta)/(1-\delta)$. Red dot-dashed lines correspond to the non-interacting limit $\delta=0$. Solid black line corresponds to the susceptibility obtained via the hydrodynamic approach, namely $f_L(\omega)=1$, while the dashed blue line includes the effect of Luttinger liquid renormalization of the spectral weight. Inset: dispersion of the Larmor and magnetization-current modes at small momentum obtained within the hydrodynamic approach \cite{KBS2020,RBWang2022}. A gap between the two branches in the dispersion results in a spectral gap in the local susceptibility for $\delta\neq0$. (c,d) Spin current as a function of temperature obtained using Eqns.\ (\ref{eq2}) and (\ref{eq8b}) and the local susceptibilities in (a,b) for $\delta > 0$ in (c) and $\delta < 0$ in (d). High-temperature limit of (c,d) is sketched in Fig.1(b).}
    \label{fig:chi}
\end{figure}

Here we illustrate field-theoretical predictions for the end-chain spin susceptibility, Eq.\ (\ref{eq7}), and the low-temperature asymptotic of the spin current, Eq.\ (\ref{eq8b}), in the main text.

Figures \ref{fig:chi}(a,b) show the analytical result Eq.\ (\ref{eq7}) for the local susceptibility at the end of the spin chain for two signs of the spinon interaction $\delta$. 

{\bf Low T limit.} As $T$ increases from zero, the effective range of $\omega$ increases with it, and for $T_{\rm sign}$ of order $h$ (numerically, we find it to be approximately $0.3 h$), the negative frequency contribution $\tilde{j}_{\rm a}$ in (\ref{eq8b}) overcomes those from $\tilde{j}_{\rm b}$ and $\tilde{j}_{\rm c}$. As a result, for $T > T_{\rm sign}$, the net spin current $j$ {\em changes sign}. This unusual behavior is illustrated in Fig.~\ref{fig:chi}(c,d) for both signs of $\delta$. We also find that the sign-changing temperature $T_{\rm sign} \approx 0.3 h$ is not sensitive to the magnitude of $\delta$, which affects it only marginally as can be seen from the comparison of black ($f_{\rm L}=1$ for all $\omega$) and blue dashed ($f_{\rm L} \sim |\omega|^{-2\delta}$ inside the ${\rm (b)}$-interval) lines in Fig.~\ref{fig:chi}(c,d).

{\bf High T limit.} Eq.\ (\ref{eq7}) shows that magnetic field $h$ sets the scale for $\omega$. Therefore at temperature $T \gg h$ equation (\ref{eq8b}) can be simplified by approximating $\sinh(\omega/2T)$ by its argument, 
\begin{align}
\tilde{j}(\delta, h, T) = -\frac{\tilde{v}}{\chi} \int_{-\infty}^\infty \, d\omega \, {\rm Im}\chi^\pm_{{\rm S}}(\omega).
\label{end:eq8c}
\end{align}
But note that Eq.\ (4) of the main text
tells us that 
\begin{align}
&\int_{-\infty}^\infty \, d\omega \,\chi^\pm_{{\rm S}}(\omega) = 2\pi \int_{-\infty}^\infty dt \, \delta(t) (-i \theta(t)) \langle [S_1^+(t), S_1^-(0)]\rangle_0 \nonumber\\
& = - 2\pi i \theta(0) \langle [S_1^+(0), S_1^-(0)]\rangle_0 = - 2\pi i \langle S^z_1 \rangle_0 = - 2\pi i m,
\end{align}
where we used $\theta(0)=1/2$. Therefore, 
\be
\tilde{j}(\delta, h, T) = 2\pi m \tilde{v}/\chi = 2\pi \tilde{v} h > 0.
\label{end:highT}
\ee
That is, at sufficiently high $T$, the spin current {\em must be positive}. Together with the low-$T$ field theory prediction of the sign change of the spin current for {\em positive} $\delta > 0$, shown in Fig.\ \ref{fig:chi}(c), this implies {\em two} sign changes of the current! Focusing on $\delta > 0$, we, therefore, predict that at the lowest temperature, the spin current starts positive, then changes sign to the negative at a low temperature of the order $0.3 h$, as Fig.\ \ref{fig:chi}(c) shows, and then at some $T \gg h$, it turns positive again, as Fig.\ \ref{fig:setup}(b) shows. 

Exactly such a double-sign-change behavior was reported but not explained in \cite{Hirobe2017}. Notice that our field-theoretical explanation of the unexpected double-sign-change behavior of the spin current does not rely on the development of the long-range magnetic order at the low $T$, which was invoked in \cite{Hirobe2017} as the reason for the sign change at low temperature.

For completeness, we note that for $\delta < 0$, the current starts out negative and then turns to the positive at low $T \sim 0.3 h$, as is seen in Fig.\ \ref{fig:chi}(d). From there on, it stays positive for all $T$ and, unlike that in the $\delta > 0$ case, does not experience the second change of sign. This, too, is illustrated in Fig.\ \ref{fig:setup}(b). 

\section{Set-up}
{\em Antiferromagnetic spin-$\frac{1}{2}$ chain.}
The Hamiltonian of an antiferromagnetic spin-$\frac{1}{2}$ chain is given by (\ref{eq9}) in the main text where $h=g\mu_B B$ an  external magnetic field along $\hat{\bf z}$-direction.
The open $N$-site long spin chain runs from $n=1$ to $n=N$. Open boundaries imply that auxiliary boundary spins at $n=0$ and $n=N+1$ must vanish \cite{affleck1992}
\begin{align}
{\mathbf S}_{0}
&=0
={\mathbf S}_{N+1}.\label{14}
\end{align}
We take $N \to \infty$ at the end of the calculation.

{\em Metal Hamiltonian.}
The Hamiltonian for the metal is simply a free fermi system
\begin{align}
H_\mathrm{M}
&=\sum_{\mathbf r}
  {\mathbf c}_{\mathbf r }^{\dag } 
\left(-\frac{\partial_{\mathbf r}^2}{2m_e}-\epsilon_F\right)
  {\mathbf c}_{\mathbf r }
- B\sum_{\mathbf r}   s_{\mathbf r}^z 
=\sum_{\mathbf k}
  {\mathbf c}_{\mathbf k}^{\dag } 
\xi_{\mathbf k } 
  {\mathbf c}_{\mathbf k},\label{13}
\end{align}
where 
$
{\mathbf c}_{\mathbf r }
=
({ c}_{\mathbf r \uparrow},  {c}_{\mathbf r \downarrow})^\mathrm{T}
$
is annihilation operator of the spin-$\alpha=\uparrow,\downarrow$ electron at point $\mathbf{r}$ in the metal, 
$
 {\mathbf s}_{\mathbf r}
=\frac{1 }{2}
  {\mathbf c}_{\mathbf r}^{\dag }{\bm \sigma}  {\mathbf c}_{\mathbf r},$
$
 {\mathbf c}_{\mathbf r}
=\frac{1}{\sqrt{N_\mathrm{M}}}\sum_{\mathbf k} e^{i\mathbf k\cdot\mathbf r} {\mathbf c}_{\mathbf k},$ 
$
\xi_{\mathbf k}
=\frac{\mathbf k^2}{2m_e}-\epsilon_F-\frac{1}{2} B\sigma^z,
$
with $N_\mathrm M$ the number of sites in the metal.

\subsection{Spin current}
The Hamiltonian describing spin exchange interaction at the interface is given by (\ref{eq1}) in the main text.
When a temperature gradient is applied across the junction between an insulating magnet and a metal, the spin current across the interface gets generated (spin Seebeck effect). We define the spin current operator as 
\begin{align}
{I}_{\rm s} 
= \frac{d }{dt} \sum_{{\bf r}}^{N_M}   s_{\bf r}^{z}
=  - i[   s^{z}_0,   H']
=\frac{1}{2} iJ' (   S_{1}^{+}  {s}_{0}^-
-   S_{1}^{-}  {s}_{0}^+),\label{3.2}
\end{align}
where we used the fact that the Hamiltonian \eqref{13} conserves total spin $s^{z}_{\rm tot}$.
Treating $   H'$ as a perturbation, the spin current $I_{\rm s}$ flowing into the metal is found by the linear response \cite{Takahashi2016} 
\begin{align}
I_\mathrm{s} (t)
&=\delta\braket{\frac{d s^{z}_{\rm tot}}{dt} } =- i\int_{-\infty}^\infty\mathrm d t'\theta(t-t')\braket{[\frac{d s^{z}_{\rm tot}}{dt} (t),   H'(t')]}_0,\label{3.4}
\end{align}
where  
\begin{align}
\braket{  {A}}_0
=\operatorname{Tr}(  {\rho}_0  {A})
=\frac{1}{Z}\sum_{\mathbf k \alpha n}\bra{\mathbf k\alpha n}e^{-\beta_\mathrm{S}  {H}_\mathrm{S}-\beta_\mathrm M  {H}_\mathrm M}  {A}\ket{\mathbf k\alpha n},
\end{align}
is the thermal average with respect to the unperturbed Hamiltonian 
$
   H_0
=   H_\mathrm{S}+   H_\mathrm{M}
$
and we have assumed the spin chain and the metal have reached their respective thermal equilibrium,
$
  {\rho}_0
=\frac{1}{Z}e^{-\beta_\mathrm{S}  {H}_{S}-\beta_\mathrm M  {H}_\mathrm{M}},
Z
=\operatorname{Tr}   {\rho}_0
$
with $\ket n$ and $\ket{\mathbf k\alpha}$ energy eigenstates of $  {H}_{\rm S}$ and $  {H}_\mathrm{M}$, respectively,
$
  {H}_{\rm S}\ket n
=E_n\ket n,
$ 
$
  {H}_\mathrm{M}\ket{\mathbf k\alpha}
=\xi_{\mathbf k\alpha}\ket{\mathbf k\alpha},$ 
$
\ket{\mathbf k\alpha}
=  {c}_{\mathbf k\alpha}^\dag\ket0,$ $
  {c}_{\mathbf k\alpha}\ket0
=0.
$

Equations \eqref{3.4} and \eqref{3.2} lead to
\begin{align}
I_\mathrm{s} (t)
&=\frac{J'^2}{2}\operatorname{Re} \int_{-\infty}^\infty
\mathrm dt\theta(t)
\left[\braket{   S_{1}^{+}(t)   S_{1}^{-}}_0
\braket{   s_{0}^{-}(t)   s_{0}^{+}}_0\right.\nonumber\\
&\left.-\braket{   S_{1}^{-}(-t)   S_{1}^{+}}_0
\braket{   s_{0}^{+}(-t)   s_{0}^{-}}_0\right],\label{3.14}
\end{align}
so the original thermal average is now decoupled into respective thermal Green's functions
\begin{align}
G_\mathrm{S}^>(t)
\equiv\braket{   S_{1}^{+}(t)   S_{1}^{-}}_0,\quad
g_\mathrm M^<(t)
\equiv \braket{   s_{0}^{-}(t)   s_{0}^{+}}_0.\label{S7}
\end{align}
The Fourier transform of these local thermal Green's functions simply relates the  imaginary part of their respective local transverse spin susceptibilities  in frequency space as 
\begin{align}
G_\mathrm {S}^>(\omega)
&=\int_{-\infty}^\infty
\mathrm dt e^{i\omega t}G_\mathrm {S}^>(t)\nonumber\\
&=-2[1+n_B(\omega,T_S)]\operatorname{Im} \chi^{+-}_\mathrm{S}(\omega),\\
g_\mathrm M^<(\omega)
&\equiv \int_{-\infty}^\infty
\mathrm dt e^{i\omega t}g_\mathrm M^<(t)\nonumber\\
&=-2n_B(-\omega,T_\mathrm M)\operatorname{Im} \chi^{+-}_\mathrm {M}(-\omega),
\end{align}
where 
$n_B(\omega,T)$
is the Bose function, and the local transverse spin susceptibilities for the spin chain and the metal are defined as 
\begin{align}
\chi^{+-}_\mathrm {S}(t)
&=-i\theta(t)\braket{[   S_{1}^{+}(t),   S_{1}^{-}]_-}_0,\label{3.21}\\
\chi^{+-}_\mathrm {M}(t)
&=-i\theta(t)\braket{[   s_{0}^{+}(t),   s_{0}^{-}]_-}_0,\label{S11}
\end{align}
respectively.
Then, from \eqref{S7}-\eqref{S11}, we find the spin current \eqref{3.14} is just a convolution of the imaginary part of the  local transverse spin susceptibilities of the spin chain and the metal with the Bose factors in frequency space \cite{Hirobe2017,Takahashi2016}.
\begin{align}
I_\mathrm{s} (t)
&=J'^2
\int_{-\infty}^\infty\frac{\mathrm d \omega }{2\pi} 
[n_B(\omega,T_\mathrm M)-n_B(\omega,T_\mathrm S)]\nonumber\\
&\times\operatorname{Im} \chi^{+-}_\mathrm {S}(\omega) 
\operatorname{Im} \chi^{+-}_\mathrm {M}(\omega).\label{3.25}
\end{align}

\subsection{Transverse spin susceptibility of the metal}
The transverse spin susceptibility of metal \eqref{S11} can be written in momentum space
\begin{align}
\chi^{+-}_\mathrm {M}(t)
&=-i\theta(t)\frac{1}{N_\mathrm M^2}\sum_{\mathbf q\mathbf k}e^{i(\xi_{\mathbf k\uparrow}-\xi_{\mathbf k+\mathbf q\downarrow})t}\nonumber\\
&\times
[n_F(\xi_{\mathbf k \uparrow})
-n_F(\xi_{\mathbf k+\mathbf q\downarrow})],
\end{align}
where $n_F(\omega)$ is the Fermi function.
Then in frequency space, it becomes
\begin{align}
\chi^{+-}_\mathrm {M}(\omega)
&\equiv \int_{-\infty}^\infty\mathrm dt e^{i\omega t} \chi^{+-}_\mathrm {M}(t)\nonumber\\
&=\int\frac{\mathrm d^3q}{(2\pi)^3}\int\frac{\mathrm d^3k}{(2\pi)^3}
\frac{n_F(\xi_{\mathbf k\uparrow})
-n_F(\xi_{\mathbf k+\mathbf q\downarrow})}{\omega+\xi_{\mathbf k\uparrow}-\xi_{\mathbf k+\mathbf q\downarrow}+i0^+}.\label{3.29}
\end{align}
Hence the imaginary part of \eqref{3.29} becomes
\begin{widetext}
\begin{align}
&\operatorname{Im}\chi^{+-}_\mathrm{M}(\omega)
=-\pi\int\frac{\mathrm d^3q}{(2\pi)^3}\int\frac{\mathrm d^3k}{(2\pi)^3}
[\theta(-\xi_{\mathbf k\uparrow})
-\theta(-\xi_{\mathbf k+\mathbf q\downarrow})] 
\delta(\omega+\xi_{\mathbf k\uparrow}-\xi_{\mathbf k+\mathbf q\downarrow})\label{app:o0}\\
&=-\pi\int\frac{\mathrm d^3q}{(2\pi)^3}\frac{1}{(2\pi)^2}\int_0^\infty k^2\mathrm dk\int_{-1}^1\mathrm d\cos\theta
[\theta(\xi_{\mathbf k\uparrow})
-\theta(-\omega-\xi_{\mathbf k\uparrow})]\delta\left(\omega- B-\frac{q^2+2kq\cos\theta}{2m}\right)\label{S16}\\
&=
\begin{cases}
-\frac{1}{4\pi}\int\frac{\mathrm d^3q}{(2\pi)^3}\int_{k_2}^{k_1}k^2\mathrm dk\int_{-1}^1
\frac{m}{kq}\delta\left(\frac{m_e}{kq}(\omega- B)-\frac{q}{2k}-\cos\theta\right)\mathrm d\cos\theta,&\quad\omega>0\\
\frac{1}{4\pi}\int\frac{\mathrm d^3q}{(2\pi)^3}\int_{k_1}^{k_2}k^2\mathrm dk\int_{-1}^1
\frac{m}{kq}\delta\left(\frac{m_e}{kq}(\omega- B)-\frac{q}{2k}-\cos\theta\right)\mathrm d\cos\theta,&\quad\omega<0
\end{cases}\label{S17}
\end{align}
\end{widetext}
where in \eqref{S16} we have used the spherical coordinates for the $k$-integration, 
and in \eqref{S17} we have used,
for given $(q,\omega)$,
\begin{align}
&\xi_{\mathbf k\uparrow}-\xi_{\mathbf k+\mathbf q\downarrow}
=-\frac{\mathbf q^2+2\mathbf k\cdot\mathbf q}{2m_e}- B,\\
&\theta(-\xi_{\mathbf k\uparrow})
=1,\quad 0\leq k\leq\sqrt{2m_e(\epsilon_F+\frac{B}{2})}\equiv k_1,\\
&\theta(-\omega-\xi_{\mathbf k\uparrow})
=1,\quad 0\leq k\leq\sqrt{2m_e(\epsilon_F+\frac{B}{2}-\omega)}\equiv k_2.
\end{align}
For small frequency and magnetic field (relative to the Fermi energy) $\omega,B\ll \epsilon_F $, 
$
k_1
\approx k_F(1+\frac{B}{4\epsilon_F})
$
and
$
k_2
\approx k_F(1+\frac{B}{4\epsilon_F}-\frac{\omega}{2\epsilon_F}),
$
and finally, \eqref{S17} gives
\begin{align}
\operatorname{Im}\chi^{+-}_\mathrm {M}(\omega)
&\approx-\frac{m_e^2k_F^2}{4\pi^3}\omega = -\pi {\cal D}_F^2 \omega.\label{4.14}
\end{align}
where ${\cal D}_F$ is the single-particle density of states at the Fermi level.

In this simple calculation, we did not take into account an open boundary condition which the metal is obviously subject to, similar to the spin chain that connects to it. However, we do not expect the answer to change when the open BC is accounted for. This follows from the fact that (a) the ratio $B/\epsilon_F \ll 1$ for the metal, and (b) ${\rm Im}\chi^{+-}_\mathrm{M}(\omega)$ in \eqref{app:o0} represents the standard result for the particle-hole continuum of the magnetized electron gas. Correction to \eqref{4.14} goes as $(\omega -B)/\epsilon_F \ll 1$.

This statement is fully supported by the independent calculation of ${\rm Im}\chi^{+-}(\omega)$ in Sec.\ \ref{app:free-open-chain} for the case of non-interacting fermion chain subject to the magnetic field {\em and} open boundary condition. The result, Eq.\ \eqref{app:12}, shows that ${\rm Im}\chi^{+-}(\omega) \sim \omega$ at low energy, and the correction to this generic behavior comes the $B/t \sim B/\epsilon_F \ll 1$ term which is essentially zero for any reasonable conductor.

\section{Transverse spin susceptibility of the chain}
\label{app:trans}

Here we sketch how to obtain the spin susceptibility of an antiferromagnetic spin-$\frac{1}{2}$ chain with open boundary conditions (OBC) \eqref{14} using the hydrodynamic technique described in \cite{RBWang2022}. We follow \cite{Fabrizio1995} to implement OBC.

\subsubsection{Low-energy description}
At low energies, the lattice spin operator is  approximated as \cite{gogolin2004bos}
\begin{align}
   {\mathbf S}_b
\to a   {\mathbf S}(x)
=a  {\bm \psi}^\dag ( x)\frac{\bm\sigma}{2}  {\bm\psi} ( x),\label{4.18}
\end{align}
where $a$ is the lattice constant, $x=ba$, and 
$
  {\bm\psi} ( x)
=\begin{pmatrix}
  {\bm\psi} _\uparrow( x)\\
  {\bm\psi} _\downarrow( x)
\end{pmatrix}
$
is the spinon operator.
The open boundary conditions \eqref{14} requires that 
\begin{align}
  \psi_\alpha ( 0)
&=0
=  \psi_\alpha (L),\label{boundcond}
\end{align}
where $\alpha=\uparrow,\downarrow$, $L=Na$ is the length of the spin chain.
The Fourier  expansion of the $  \psi$-operator, appropriate for the boundary conditions (\ref{boundcond}), takes the form \cite{Fabrizio1995}
\begin{align}
  \psi_\alpha (x)
&=\sqrt{\frac{2}{L}}\sum_{n=1}^\infty \sin (k_nx)   c_{\alpha k_n}
=-  \psi_\alpha (-x)
=  \psi_\alpha (x+2L),\label{S24}
\end{align}
where 
$
k_n
=\frac{n\pi}{L}.
$
As a result, the standard slow-varying Fermi fields $  \psi_{\alpha R/L} (x)$
\begin{align}
  \psi_\alpha (x)
&\equiv e^{ik_F x}  \psi_{\alpha R} (x) +  e^{-ik_F x}  \psi_{\alpha L} (x) ,\label{decomposition}
\end{align}
where $ k_F=\frac{n_F\pi}{L}$ is the Fermi momentum, from \eqref{S24}, are found to be connected as follows 
\begin{align}
\psi_{\alpha R} (x)
&=  \psi_{\alpha R} (x+2L) \label{4.25}\\
&=-  \psi_{\alpha L}(-x)\label{4.26}.
\end{align}
This leads to the  spin density \eqref{4.18} parameterization
\begin{align}
   {\mathbf S}(x)
=   {\mathbf J}_L(x)+   {\mathbf J}_R (x) + (-1)^{\frac{x}{a}}    {\mathbf N}(x),\label{4.27}
\end{align}
where at half filling $2n_F=N$,
\begin{align}
{\mathbf J}_r(x)
=
\substack{\boldsymbol{\cdot}\\\boldsymbol{\cdot}}
   {\bm\psi}_{ r}^\dag(x)\frac{\bm\sigma}{2}  {\bm\psi}_{ r}(x)
\substack{\boldsymbol{\cdot}\\\boldsymbol{\cdot}}\label{S29}
\end{align}
is the uniform components of the $r=R,L$-moving spin-currents and 
\begin{align}
{\mathbf N} (x)
=
\substack{\boldsymbol{\cdot}\\\boldsymbol{\cdot}}
   {\bm\psi}_{ R}^\dag(x)\frac{\bm\sigma}{2}  {\bm\psi}_{L}(x)
+   {\bm\psi}_{L}^\dag(x)\frac{\bm\sigma}{2}  {\bm\psi}_{R}(x)
\substack{\boldsymbol{\cdot}\\\boldsymbol{\cdot}}
\label{eqN}
\end{align}
is the staggered component of the local spin density. 
Together with \eqref{4.26} and \eqref{S29}, the staggered component \eqref{eqN} 
near the open boundary reduces to
\begin{align}
  {\mathbf N} (x)
&=
\substack{\boldsymbol{\cdot}\\\boldsymbol{\cdot}}
-   {\bm\psi}_{ R}^\dag(x)\frac{\bm\sigma}{2}  {\bm\psi}_{R}(-x)
-   {\bm\psi}_{R}^\dag(-x)\frac{\bm\sigma}{2}  {\bm\psi}_{R}(x)
\substack{\boldsymbol{\cdot}\\\boldsymbol{\cdot}} \to \nonumber\\
&\to - 2   {\mathbf J}_R (0) \, {\rm for} \, x\approx 0.
\label{eqN2}
\end{align}
showing its relation to the uniform component at the boundary.
Note that the chiral spin currents are not independent and are periodic to twice  the chain length $L$ due to  \eqref{4.25} and \eqref{4.26}
\begin{align}
  {\mathbf J}_R(x)
&=  {\mathbf J}_L(-x)\label{4.30}\\
&=  {\mathbf J}_R(x+2L),\label{4.31}
\end{align}
and they  obey the Kac-Moody algebra 
\cite{gogolin2004bos}
\begin{align}
[   J_{R}^a(x),   J_{R}^b(x')] 
&=i[\frac{-1}{4\pi} \delta'(x-x')\delta^{ab} 
+ \delta(x-x')  \epsilon^{abc}    J_{R}^c(x) ],\label{4.24}
\end{align} 
where the prime symbol on the first delta function denotes the derivative with respect to its argument. This is the key element of our hydrodynamic technique \cite{RBWang2022},

One can actually now work with the right-moving field only and write the low energy effective Hamiltonian in the Sugawara form, which is quadratic in the $SU(2)$ spin densities \cite{gogolin2004bos}, and  unfold it onto $(-L, L)$ interval \cite{Fabrizio1995} due to \eqref{4.30} and \eqref{4.31} as
\begin{align}
   H_{\rm S}
=   H_{\rm free}+   H_\mathrm{ bs}+   V.\label{eq:h_nab}
\end{align}
Here 
\begin{align}
   H_{\rm free}
&=\frac{2\pi v}{3}\int_{0}^L \mathrm dx
\substack{\boldsymbol{\cdot}\\\boldsymbol{\cdot}}
   {\mathbf J}_R(x)\cdot   {\mathbf J}_R(x)+   {\mathbf J}_L(x)\cdot   {\mathbf J}_L(x)
\substack{\boldsymbol{\cdot}\\\boldsymbol{\cdot}}\\
&=\frac{2\pi v}{3}\int_{-L}^L \mathrm dx
\substack{\boldsymbol{\cdot}\\\boldsymbol{\cdot}}
   {\mathbf J}_R(x)\cdot   {\mathbf J}_R(x)
\substack{\boldsymbol{\cdot}\\\boldsymbol{\cdot}}
\label{eq:h_def}
\end{align}
is the spin part of the non-interacting Hamiltonian of Dirac fermions, and 
$
v
\approx\frac{1}{2} J a \pi
$
is the spinon speed. 
External Zeeman field enters via 
\begin{align}
   V
&=-h\int_{0}^L \mathrm dx [   J_R^z(x)+    J_L^z(x)]
=-h\int_{-L}^L \mathrm dx    J_R^z(x) .\label{2.82}
\end{align}
Interaction between spinons is described by the backscattering term
\begin{align}
   H_\mathrm{ bs}
&=-g_\mathrm{ bs}\int_0^L \mathrm dx
\substack{\boldsymbol{\cdot}\\\boldsymbol{\cdot}}
  {\mathbf J}_R(x)\cdot    {\mathbf J}_L(x)
\substack{\boldsymbol{\cdot}\\\boldsymbol{\cdot}}\nonumber\\
&=-\frac{g_\mathrm{ bs}}{2}\int_{-L}^L \mathrm dx
\substack{\boldsymbol{\cdot}\\\boldsymbol{\cdot}}
  {\mathbf J}_R(x)\cdot    {\mathbf J}_R(-x)
\substack{\boldsymbol{\cdot}\\\boldsymbol{\cdot}},
\label{2.81}
\end{align}
which is the leading marginally irrelevant interaction with backscattering strength $g_\mathrm{ bs}$. 

Eq.\ \eqref{eq:h_nab} describes $J_1$-$J_2$ Heisenberg chain in the vicinity of the critical end-point of the Luttinger phase located at $g_{\rm bs}=0$. 
In terms of the lattice spin model this corresponds to $J_{2,c} \approx 0.241 J_1$ \cite{eggert1996,KBS2020}.

\subsubsection{Equations of motion}

To construct the hydrodynamic equations, we first define the magnetization and magnetization current as 
\begin{align}
  {\mathbf M}(x,t)
&=  {\mathbf J}_{R}(x,t)+  {\mathbf J}_{R}(-x,t),\\
  {\mathbf J}(x,t)
&=  {\mathbf J}_{R}(x,t)-  {\mathbf J}_{R}(-x,t).
\end{align}
Equations of motion for these fields, using Hamiltonian \eqref{eq:h_nab} and the Kac-Moody algebra \eqref{4.24}, are derived to be \cite{RBWang2022}
\begin{align}
\partial_t  {\mathbf M}(x,t)
&=-v(1+\delta)\partial_x  {\mathbf J}(x,t)
-h\hat{\mathbf z}\times  {\mathbf M}(x,t),\label{4.32}\\
\partial_t  {\mathbf J}(x,t)
&=-v(1-\delta)\partial_x  {\mathbf M}(x,t)\nonumber\\
&-g_\mathrm{bs}   {\mathbf M}(x,t)\times  {\mathbf J}(x,t)
-h\hat{\mathbf z}\times  {\mathbf J}(x,t),\label{4.33}
\end{align}
where 
$
\delta
=\frac{1}{2}g_\mathrm{bs}\chi_0,
$
is the dimensionless interaction parameter and 
$
\chi_0
=\frac{1}{2\pi v}
$
is the non-interacting spinon susceptibility.

According to \eqref{4.27} and \eqref{eqN2}, the spin density at $x=0$ vanishes,
$   {\mathbf S}(0)
=  {\mathbf M}(0) +     {\mathbf N}(0)
=0,
$
therefore, at the left-most end of the chain $x=a$ we obtain 
\begin{align}
{\mathbf S}(a)
&=  {\mathbf M}(a) -     {\mathbf N}(a)
\approx   {\mathbf M}(0) -     {\mathbf N}(0)
= 2  {\mathbf M}(0) , \label{95}
\end{align} 
up to irrelevant terms with spatial derivatives.

This important result tells us that dynamical spin susceptibility at the open end of the spin chain reduces to the dynamical uniform susceptibility and is fully expressed in terms of the spin current operators $  {\mathbf J}_R(x,t)$ and $  {\mathbf J}_R(-x,t)$, in agreement with the classic paper by Affleck and Eggert \cite{affleck1992}.

\subsubsection{Equilibrium magnetization}
To find the expectation value ${\bf m}$ of the magnetization, 
\begin{align}
\mathbf m
=\braket{   {\mathbf M}(x)},
\end{align}
we resort to the mean-field treatment of the backscattering term \eqref{2.81}, 
\begin{align}
   H_\mathrm{ bs}
&=-\frac{1}{2}g_\mathrm{ bs}m\int_{-L}^L {\rm d}x  { J}_R^z(x). \label{eq:BSmf}
\end{align}
Hence the total field experienced by spinons  is given by the sum of \eqref{2.82} and \eqref{eq:BSmf}
\begin{align}
   V+   H_\mathrm{ bs}
&\approx-(h+\frac{1}{2}g_\mathrm{ bs}m)\int_{-L}^L \mathrm dx    J_R^z(x).
\end{align}
The self-consistency equation 
\begin{align}
\chi_0(h+\frac{1}{2}g_\mathrm{ bs}m)
&=m,
\end{align}
produces the desired equilibrium magnetization 
\begin{align}
m
=\chi h, 
\end{align}
where $\chi=\frac{\chi_0}{1-\delta}$
is the renormalized spinon susceptibility.

\subsubsection{Hydrodynamic equations}
We can now linearize the equations of motion \eqref{4.32} and \eqref{4.33} by considering the quantum fluctuation operators
$
\delta    {\mathbf m}(x,t)
\equiv   {\mathbf M}(x,t)-m  \hat {\mathbf z}$
and 
$
\delta    {\mathbf j}(x,t)
\equiv  {\mathbf J}(x,t),
$
to obtain the linearized hydrodynamic equations 
\begin{align}
(\partial_t+h  \hat {\mathbf z}\times)\delta    {\mathbf m}(x,t)
&=-v(1+\delta)\partial_x \delta    {\mathbf j}(x,t),\\
(\partial_t+\frac{1+\delta}{1-\delta}h  \hat {\mathbf z}\times)\delta    {\mathbf j}(x,t)
&=-v(1-\delta)\partial_x\delta    {\mathbf m}(x,t).
\end{align}

In Fourier space, we find opposite circular components decouple from each other, 
\begin{align}
(\omega\mp h)\delta    m_\pm
&=(1+\delta)vk\delta    j_\pm,\\
(\omega\mp h')\delta    j_\pm
&=(1-\delta)vk\delta    m_\pm,
\end{align}
where $h $ is the Zeeman frequency  and $h'=\frac{1+\delta}{1-\delta}h$ is the precession frequency of the spin-current mode, as well as from the longitudinal fluctuations
\begin{align}
\omega\delta    m_z
&=(1+\delta)vk\delta    j_z,\\
\omega\delta    j_z
&=(1-\delta)vk\delta    m_z.
\end{align}

\subsubsection{Green's functions}
Define the retarded Green's function matrix as 
\begin{align}
&G^{ab}(x,t;x',t')
=-i\theta(t-t')\braket{[   \psi^a(x,t),   \psi^b(x',t')]_-}\\
&=\int_{-\infty}^\infty\frac{\mathrm dk}{2\pi} e^{ik(x-x')}
\int_{-\infty}^\infty\frac{\mathrm d\omega}{2\pi}e^{-i\omega(t-t')}G^{ab}(k,\omega),
\end{align} 
where 
${\bm\psi}(x,t)
=
(
   M^+,
   M^-,
   M^z,
   J^+,
   J^-,
   J^z
)^\mathrm{T},
$
Then its equation of motion becomes
\begin{align}
\partial_t G^{ab}(x,t;x',t')
&=-i\delta(t-t')\braket{[   \psi^a(x,t),   \psi^b(x',t)]_-}\nonumber \\
&-i\theta(t-t')\braket{[\partial_t\delta    \psi^a(x,t),\delta    \psi^b(x',t')]_-},\label{S58}
\end{align}
where 
$
\bm\delta   {\bm\psi}(x,t)
=
(
\delta    m^+,
\delta    m^-,
\delta    m^z,
\delta    j^+,
\delta    j^-,
\delta    j^z
)^\mathrm{T}.
$
Using the Kac-Moody algebra \eqref{4.24} in \eqref{S58}, we find the transverse spin susceptibility for spinons in sought is then obtained by solving the matrix equation in Fourier space
\begin{align}
\omega
\begin{pmatrix}
G^{12}\\
G^{42}
\end{pmatrix}
=\begin{pmatrix}
2m\\
\frac{k}{\pi}
\end{pmatrix}
+\begin{pmatrix}
h&(1+\delta)vk\\
(1-\delta)vk&h'\\
\end{pmatrix}
\begin{pmatrix}
G^{12}\\
G^{42}
\end{pmatrix},
\end{align}
where the frequency is understood to have an infinitesimally small imaginary part $\omega\to\omega+i0^+$. The transverse dynamical spin susceptibility is found to be
\begin{align}
&\chi^{+-}_\mathrm S(k,\omega) = G^{12}(k,\omega\to\omega+i0^+)= \nonumber\\
&=\chi_0[\frac{A_+(k)}{\omega-\omega_{+}(k)+i0^+}
+\frac{A_-(k)}{\omega-\omega_{-}(k)+i0^+}].\label{4.57}
\end{align}
The excitation energies of hybridized Larmor and spin-current modes are 
\begin{align}
\omega_{\pm}(k)
&=\frac{h}{1-\delta}
\pm\sqrt{(\frac{\delta}{1-\delta}h)^2+(\tilde{v} k)^2},\label{7.4}
\end{align}
and the corresponding spectral weights
\begin{align}
A_\pm(k)
&=\frac{h}{1-\delta}\pm\frac{-\delta(\frac{h}{1-\delta})^2+(1+\delta)(vk)^2}{\sqrt{(\frac{\delta }{1-\delta}h)^2+(\tilde{v} k)^2}},\label{7.30}
\end{align}
and 
$
\tilde{v}
=\sqrt{1-\delta^2}v
$
is the renormalized spinon velocity.

It is worth noting that our result \eqref{4.57}-\eqref{7.30}  
coincides with the expression obtained earlier for the uniform infinite spin chain in papers \cite{KBS2020,RBWang2022}. Yet, as our derivation here shows, it accounts fully for the open boundary at $x=0$, which is crucial for the physics of the current problem.

\subsubsection{Local susceptibility of the AF spin-$\frac{1}{2}$ chain}
The local susceptibility of the AF spin-$\frac{1}{2}$ chain is found by integrating over the momentum in \eqref{4.57}. 
Its imaginary part is given by
\begin{widetext}
\begin{align}
\operatorname{Im}\chi^{+-}_\mathrm {S}(\omega)
&=-\pi\chi_0\int_{-\infty}^\infty\frac{\mathrm dk}{2\pi}
\left[A_+(k)\delta\left(\omega-\omega_{+}(k)\right)
+A_-(k)\delta\left(\omega-\omega_{-}(k)\right)\right]\nonumber\\
&=
- \frac{\chi \,\omega}{\tilde{v}} 
  \begin{cases}
 \sqrt{\frac{h'-\omega}{h-\omega}}, & \, \omega < -h_{\rm min};\\
  \sqrt{\frac{h'-\omega}{h-\omega}}, & \, -h_{\rm min} <\omega < h_{\rm min};\\
  0, & \, h_{\rm min} < \omega < h_{\rm max}; \\
  \sqrt{\frac{\omega-h'}{\omega-h}}, & \omega > h_{\rm max} ; 
\end{cases}\label{4.61}
\end{align}
\end{widetext}
where we have defined $h_{\rm min} = {\rm min}(h, h')$ and $h_{\rm max} = {\rm max}(h, h')$. In this way, $h_{\rm min} \, (h_{\rm max}) = h \, (h')$ for $\delta > 0$, correspondingly, while for negative $\delta < 0$ they switch and $h_{\rm min} \, (h_{\rm max}) = h' \, (h)$. As written, Eq.\ \eqref{4.61} holds for any sign of $\delta$.

We see that the $k=0$ spectral gap with the Larmor and spin-current modes in \eqref{7.4} translates into the vanishing spectral weight in the interval $h_{\rm min} < \omega < h_{\rm max}$ in \eqref{4.61}.

For the non-interacting case  $\delta\to0$ the spectral gap vanishes, $h'\to h$, and
\begin{align}
\operatorname{Im}\chi^{+-}_\mathrm {S}(\omega)
&\to -\chi_0\frac{\omega}{v}, \label{app:d0}
\end{align}
which is odd in frequency $\omega$ and leads to zero spin current.

Eq.\ \eqref{4.61} is the key analytical result of this work.

\subsubsection{Spin current from an antiferromagnetic spin-$\frac{1}{2}$ chain}

From \eqref{3.25}, the spinon spin current reads
\begin{align}
I_{\rm s} 
&= J'^2 \int_{-\infty}^\infty \frac{d \omega}{2\pi} \Big( n_B(\omega, T_{\rm M}) - n_B(\omega, T_{\rm S}) \Big) \nonumber\\ 
&\times {\rm Im}\chi^\pm_{{\rm M}}(\omega) \, {\rm Im}\chi^\pm_{{\rm S}}(\omega) .
\label{app:eq4}
\end{align}
When the temperature difference between the spin chain and metal is small, $T_{\rm M} - T_{\rm S} = \Delta T \ll T = (T_{\rm S} + T_{\rm M})/2$, the difference of Bose functions simplifies to $\omega \Delta T/(4 T^2 \sinh^2[\omega/2T])$.
This leads to the equation (3) in the main text.

Next, using \eqref{4.14} and \eqref{4.61} we obtain $I_{\rm s} = C \,\tilde{j} \,\Delta T$ and recover Eq.\ (7) in the main text,
\be
\tilde{j}(\delta, h, T) = -\frac{\tilde{v}}{\chi} \int_{-\infty}^\infty \, d\omega \frac{\omega^2 \, {\rm Im}\chi^\pm_{{\rm S}}(\omega)}{4 T^2 \sinh^2(\frac{\omega}{2T})} ,
\label{app:eq8b}
\ee
where $C = J'^2 {\cal D}_F^2 \chi/(2 \tilde{v})$. This form clearly shows that the current is determined by the not-odd-in-$\omega$ terms in the local susceptibility of the open spin chain and vanishes for $\delta =0$ when \eqref{app:d0} applies.

\subsection{Low-energy Luttinger physics}

At energies much smaller than the Zeeman energy $h$ Hamiltonian \eqref{eq:h_nab} reduces to that of non-interacting {\em bosons}. This well-understood fact \cite{affleckoshikawa1999,gogolin2004bos,giamarchi2003} is a consequence of the strongly oscillating (in space) nature of the transverse part of the backscattering term $\propto J^+_R(x) J^{-}_R(-x)$ which has the effect of eliminating this term from the Hamiltonian. As a result, at energies well below $h$ the effective Hamiltonian is given by \eqref{eq:h_nab} but with the backscattering term given by 
\begin{align}
H_{\rm bs} = -\frac{g_{{\rm bs},z}}{2} \int_{-L}^L dx \, J^z_R(x)J^z_R(-x)
\label{app:BSlow}
\end{align}
The free part reads
\begin{align}
H_{\rm free} = 2\pi v \int_{-L}^L dx \, J^z_R(x) J^z_R(x)
\label{app:H0low}
\end{align}
Together, these equations describe an XY-like model with $U(1)$ symmetry. They are conveniently analyzed with the help of abelian bosonization \cite{Fabrizio1995}. Namely, switching to the description at constant magnetization $m$ we have 
\begin{align}
J^z_R(x) = m + \frac{1}{\sqrt{2\pi}} \partial_x \Phi_R(x),
\label{app:1}
\end{align}
where $\Phi_R$ is expanded in boson modes $   b_n$, with $q_n = \pi n/L$,
\begin{align}
\Phi_R(x) = \sum_{n=1}^\infty \frac{e^{-a q_n/2}}{\sqrt{4\pi n}} \Big( e^{i q_n x}   {b}_n + e^{-i q_n x}   {b}^\dagger_n \Big).
\label{app:2}
\end{align}
The total Hamiltonian is quadratic but not diagonal in bosons $   b_n$. It is diagonalized by 
\be
\Phi_R(x) = c \varphi(x) + s \varphi(-x), \label{app:3}
\ee
subject to $c^2 - s^2=1$, which is solved by $c=\cosh\theta, s=\sinh\theta$. Here field $\varphi(x)$ has mode expansion similar to $\Phi_R(x)$ but in terms of new bosons $\hat{a}_n, \hat{a}^\dagger_n$. Diagonalization requires $\tanh(2\theta)=-g_{{\rm bs},z}/(4\pi v)$ and results in 
\begin{align}
\hat H_{\rm free} + \hat H_{\rm bs} 
= \sum_n e^{-a q_n} \frac{v q_n}{\cosh(2\theta)} a_n^\dagger a_n
\end{align}
Here $v/\cosh(2\theta) = v'$ is the new velocity. 
To calculate the local transverse spin susceptibility at the edge we need 
\begin{align}
J_R^+(0,t) \sim e^{-i \sqrt{8\pi}\Phi_R(0,t)} = e^{-i \sqrt{8\pi} e^{\theta}\varphi(0,t)},
\label{app:4}
\end{align}
where the last relation follows from \eqref{app:3}. Observe that the Luttinger constant $K = e^{-2\theta} \approx 1 + \delta$, where we use that $g_{{\rm bs},z} \approx g_{\rm bs}(h)$ according to the RG flow \cite{affleckoshikawa1999}.
The susceptibility is given by 
\begin{align}
\chi_{\rm S}^\pm(t) 
= -i \theta(t) \langle S^+(t) S^-(0) - S^-(0) S^+(t) \rangle 
\equiv -i \theta(t) \Phi(t),
\end{align}
where $S^+(t) = 4 J^+_R(0,t)$.
Using spectral decomposition, it is easy to show that the first (second) term determines ${\rm Im} \chi^\pm(\omega)$ for $\omega > 0$ ($\omega <0$), correspondingly.  With the help of Baker-Hausdorff identity one finds
\begin{align}
C^{+-}_0(t) &= \langle J_R^+(0,t) J^{-}_R(0,0)\rangle = \Big(\frac{a}{a+i v' t}\Big)^{2/K},\\
C^{-+}_0(-t) &= \langle J_R^-(0,0) J^{+}_R(0,t)\rangle = \Big(\frac{a}{a-i v' t}\Big)^{2/K}\nonumber
\label{app:5}
\end{align}
Therefore $\Phi(t) = - \Phi(-t) = - \Phi^*(t)$. 
The susceptibility  
\begin{align}
\chi_{\rm S}^\pm(\omega) 
= - i \int_0^\infty \, dt \, e^{i \omega t} \Phi(t)
\end{align}
obeys 
$\chi_{\rm S}^\pm(-\omega) 
= i \int_{-\infty}^0 dt e^{i \omega t} \Phi(t) 
= (\chi_{\rm S}^\pm(\omega))^*$ for real $\omega$. Hence, 
\begin{align}
\chi_{\rm S}^\pm(\omega) - (\chi_{\rm S}^\pm(\omega))^* 
= -i \int_{-\infty}^\infty \, dt \, e^{i \omega t} \Phi(t)
\end{align}
and, finally,
\begin{align}
{\rm Im}\chi_{\rm S}^\pm(\omega) 
&= -\frac{1}{2} \int_{-\infty}^\infty \, dt \, (e^{i\omega t} - e^{-i\omega t}) \Big(\frac{a}{a+ i v' t}\Big)^{2/K} \nonumber\\
&= -\frac{\pi (a/v')^{2(1-\delta)}}{\Gamma(2-2\delta)} \omega  |\omega|^{-2\delta},
\label{app:6}
\end{align}
where we used integral representation of the reciprocal gamma function \url{https://en.wikipedia.org/wiki/Reciprocal_gamma_function}.
The exponent $2/K -1 \approx 1 - 2\delta$ represents the enhancement/suppression (for $\delta > 0/\delta < 0$) of the spectral density at low frequencies by fluctuating gapless modes of the spin chain. 

This explains the function $f_{\rm L}(\omega)$ in Eq.\ (\ref{eq7}) in the main text.

\section{Lesson from the free fermion chain with open boundary}
\label{app:free-open-chain}

To understand effects due to the the finite bandwidth of the lattice model as well as those due to the curvature of the fermion dispersion, we analyzed a simple 
toy model of non-interacting fermions with open boundaries and at half-filling. 
\begin{align}
H = - \frac{t}{2} \sum_{n=1}^N (c_{s,n}^\dagger c_{s,n+1} + c_{s,n+1}^\dagger c_{s,n}) - \frac{s}{2}B \sum_n c_{s,n}^\dagger c_{s,n},
\label{app:7}
\end{align}
where $s=\pm 1$ is the spin index. Open boundaries require $c_{s,n} = \sqrt{2/N} \sum_k \sin(k n) c_{s,k}$. Then
\begin{align}
H = \sum_k ( - t \cos k - \frac{s}{2}B) c_{s,k}^\dagger c_{s,k}
\label{app:8}
\end{align}
and $S_n^+(t) = (2/N)\sum_{k,k'} \sin(k n) \sin(k' n) c_{+,k}^\dagger(t) c_{-,k'}(t)$, where the dynamics comes from $c_{s,k}(t) = e^{-i \epsilon_s(k)t} c_{s,k}$. We denote $\epsilon_s(k) = \epsilon(k) - s B/2$ and $\epsilon(k) = - t \cos k$. For the transverse susceptibility at site $n$ we find
\begin{align}
&-{\rm Im}\chi_n ^\pm(\omega) = \frac{4}{\pi} \int_0^\pi dk \int_0^\pi dk' \sin^2(k n) \sin^2(k' n) \nonumber\\ 
&\times \Big(f(\epsilon(k) - B/2) - f(\epsilon(k') + B/2)\Big) \nonumber\\
&\times \delta(\omega - B + \epsilon(k) - \epsilon(k'))
\label{app:9}
\end{align}
We consider the first site of the chain, $n=1$, and manipulate the expression into a simpler form for $B > 0$. We find that for $\omega \in (0, t + B/2)$
\begin{align}
&{\rm Im}\chi _1^\pm(\omega) = \frac{-4}{\pi t} \int_{-B/(2t)}^{-B/(2t) + \omega/t} dx \, \sqrt{1-x^2} \nonumber\\
&\times \sqrt{1- (x + (B-\omega)/t)^2},
\label{app:10}
\end{align}
while for $\omega \in (t+B/2, 2t + B)$ the integration limits are different,
\begin{align}
&{\rm Im}\chi _1^\pm(\omega) = \frac{-4}{\pi t} \int_{-1-B/t + \omega/t}^{1} dx \, \sqrt{1-x^2} \nonumber\\
&\times \sqrt{1- (x + (B-\omega)/t)^2}.
\label{app:11}
\end{align}
Expression for the negative frequencies is obtained with the help of Onsager's relation ${\rm Im}\chi^\pm(\omega, B) = - {\rm Im}\chi^\pm(-\omega, -B)$.

These results allow us to see the frequency `content' of $\chi ^\pm$. For example, in the central frequency region, it is possible to expand the susceptibility in series,
\bea
&&{\rm Im}\chi_1 ^\pm(\omega) 
= \frac{-4}{\pi t} \int_{-B/(2t)}^{-B/(2t) + \omega/t} dx \, \sqrt{1-x^2} \nonumber \\
&& \times \sqrt{1- (x + (B-\omega)/t)^2} \approx \frac{-4}{\pi t} \Big( (1-(B/2t)^2) \omega/t + \nonumber \\
&&+\frac{B}{2t} (\frac{\omega}{t})^2 - \frac{(8 - (B/t)^2)}{6(4-(B/t)^2)} (\frac{\omega}{t})^3 + \cdots\Big).
\label{app:12}
\eea
This expansion is found to work very well up to $\omega \sim t$.

\begin{figure}[t]
    \centering
    \includegraphics[width=0.85\columnwidth]{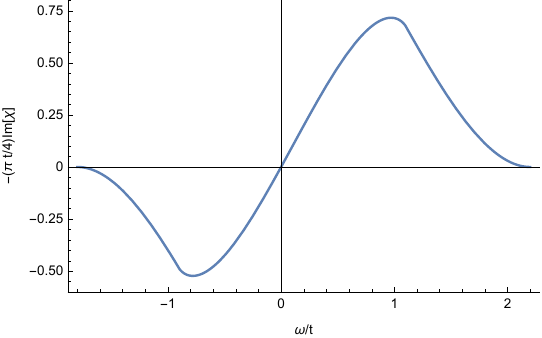}
    \caption{Transverse susceptibility $-{\rm Im}\chi^\pm_1(\omega)$ of the non-interacting chain of fermions. $B=0.2 t$.}
    \label{fig:app2}
\end{figure}

It is worth noting that the
second, $B \omega^2$, term of the expansion actually satisfies Onsager's relation $\rm {Im}\chi^\pm(\omega, B) = -{\rm Im} \chi^\pm(-\omega, -B)$. 
It is this term, {\em even} in frequency, that is responsible for the generation of the finite current $I_s(T) \sim T^3$ in (7) in the main text, if we use \eqref{app:12} in place of $\chi^\pm_{{\rm S}}$ in (6) in the main text. 

This effect is generic and shows that the non-linear dispersion of magnetic excitations is the source of the spin current at elevated temperatures.
Something like this  must also be present in the chain susceptibility at the non-interacting spinon point $J_2 = 0.24 J_1 $.

\section{Numerical simulations of the local susceptibility at finite temperatures}\label{Finte_T_calculations}

In our numerical matrix product state (MPS) simulations of the local susceptibility at finite temperatures we employ the minimally entangled typical thermal states (METTS) algorithm~\cite{White2009, Stoudenmire2010}.
The thermal expectation value of any arbitrary observable $\mathcal{O}$ can be written as 
\begin{align}
    \langle\mathcal{O}\rangle_\beta &= \frac{1}{{Z}}\sum_i\langle i|e^{-\beta H/2}\mathcal{O}e^{-\beta H/2}|i\rangle\nonumber\\
    &=\frac{1}{{Z}}\sum_i p_i\langle \psi_i|\mathcal{O}|\psi_i\rangle\label{eq:METTS_sampling},
\end{align}
where $\beta = 1/k_BT$ is the inverse temperature with $k_B=1$, ${Z}$ is the partition function, and the summation is over all the basis states. States $|\psi_i\rangle$ above are referred to as METTS states and defined as
\begin{equation}
    |\psi_i\rangle = \frac{1}{\sqrt{p_i}} e^{-\beta H/2}|i\rangle,
\end{equation}
with the probabilities $p_i=\langle i|e^{-\beta H}|i\rangle$ ensuring normalization. The METTS algorithm evaluates thermal expectation values by sampling over a \textit{finite} set of \textit{pure} METTS states.

\begin{figure}[t]
    \centering
    \includegraphics[width=0.85\columnwidth]{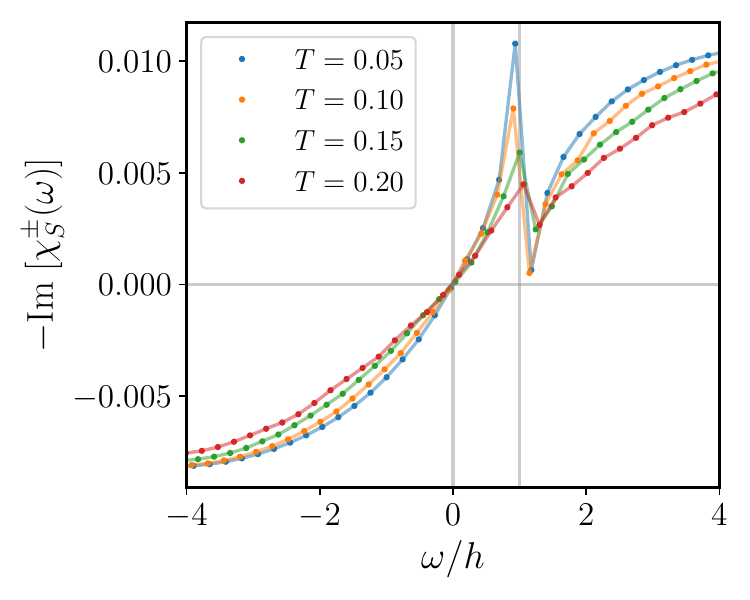}
    \caption{Transverse local susceptibility at the end of the spin chain, $-{\rm Im}\chi^\pm_{\rm{S}}(\omega)$, for $h=0.2$ at four different temperatures $T=0.05,0.10,0.15,0.20$ (we use $k_B=1$), obtained numerically using METTS simulations.}
    \label{fig:Susceptibility_vs_beta}
\end{figure}

The time-dependent correlation function $\mathcal{C}^{\pm}_\beta(t)=\langle[S_1^+(t), S_1^-(0)]\rangle_\beta$ in Eq.~(4) in the main text for the local transverse spin susceptibility is evaluated using the Dynamical METTS algorithm~\cite{Dynamical_METTS}. 
Note that Eq.~\eqref{eq:METTS_sampling} can be generalized for the case of a time-dependent correlation function as follows
\begin{equation}
    C_{\beta}^{\pm}(t)=\frac{1}{Z}\sum_ip_i\langle\psi_i|[S_1^+(t), S_1^-(0)]|\psi_i\rangle .
    \label{eq:C_beta}
\end{equation}
Denoting
\begin{equation}
    |u_i\rangle =S^-_1(0)|\psi_i\rangle, \quad
    |v_i\rangle =S^+_1(0)|\psi_i\rangle, 
\end{equation}
Eq.~\eqref{eq:C_beta} can be expressed as 
\begin{equation}
    C_{\beta}^{\pm}(t)= \frac{1}{Z}\sum_ip_i[\langle \psi_i(t)|S_1^+|u_i(t)\rangle - \langle v_i(t) |S_1^+|\psi_i(t)\rangle].
    \label{eq:Cpm_METTS}
\end{equation}

Thus to obtain $C_{\beta}^{\pm}(t)$ numerically we generated a finite set of METTS states, and carried out the time evolution of $|\psi_i\rangle, |u_i\rangle, |v_i\rangle$ using the standard time-evolving block-decimation (TEBD) method~\cite{Paeckel2019} for each METTS sample $i$. 
The number of METTS states used in the sampling was 200 for $T=0.05$, 400 for $T=0.1$, and 600 for $T= 0.15,0.2$. 
Following the time evolution up to times $T_{\rm{sim}}=65 J^{-1}$ we extrapolated $\mathcal{C}^{\pm}_\beta(t)$ up to times $2 T_{\rm{sim}}$ using linear prediction~\cite{White2008}. 
A Gaussian windowing function with standard deviation of $\sigma = T_{\rm{sim}}/\sqrt{2}$ was then applied before taking the Fourier transform to obtain $\chi^{\pm}_{\rm S}(\omega)$.
The METTS simulations were done on finite systems of length $N=120$ sites.

Results for the local susceptibility at finite temperature are presented in Fig.~\ref{fig:Susceptibility_vs_beta} showing the presence of the spectral gap in the local susceptibility up to temperatures of order the Zeeman energy $h$. 

\end{document}